# Estimating global article processing charges paid to six publishers for open access between 2019 and 2023

Stefanie Haustein[1, 2,3,*], Eric Schares[2,4], Juan Pablo Alperin[2,5], Madelaine Hare[1,2], Leigh-Ann Butler[2,6] & Nina Schönfelder[7]

[1] School of Information Studies, University of Ottawa, Ottawa (Canada)
[2] Scholarly Communications Lab, Ottawa/Vancouver (Canada)
[3] Centre interuniversitaire de recherche sur la science et la technologie (CIRST), Université du Québec à Montréal, Montréal (Canada)
[4] University Library, Iowa State University, Ames (USA)
[5] Simon Fraser University, Vancouver (Canada)
[6] University of Ottawa Library, Ottawa (Canada)
[7] Bielefeld University Library, Bielefeld (Germany)

[*] *Corresponding author:* stefanie.haustein@uottawa.ca

# Abstract

This study presents estimates of the global expenditure on article processing charges (APCs) paid to six publishers for open access between 2019 and 2023. APCs are fees charged for publishing in some fully open access journals (gold) and in subscription journals to make individual articles open access (hybrid). There is currently no way to systematically track institutional, national or global expenses for open access publishing due to a lack of transparency in APC prices, what articles they are paid for, or who pays them. We therefore curated and used an open dataset of annual APC list prices from Elsevier, Frontiers, MDPI, PLOS, Springer Nature, and Wiley in combination with the number of open access articles from these publishers indexed by OpenAlex to estimate that, globally, a total of $8.349 billion ($8.968 billion in 2023 US dollars) were spent on APCs between 2019 and 2023. We estimate that in 2023 MDPI ($681.6 million), Elsevier ($582.8 million) and Springer Nature ($546.6) generated the most revenue with APCs. After adjusting for inflation, we also show that annual spending almost tripled from $910.3 million in 2019 to $2.538 billion in 2023, that hybrid exceed gold fees, and that the median APCs paid are higher than the median listed fees for both gold and hybrid. Our approach addresses major limitations in previous efforts to estimate APCs paid and offers much needed insight into an otherwise opaque aspect of the business of scholarly publishing. We call upon publishers to be more transparent about OA fees.

# 1. Introduction

In the twenty years since the advent of the open access (OA) movement, a myriad of publishing models has emerged as alternatives to well-established subscription models, contributing to the growing complexity of the OA publishing landscape. One increasingly popular approach to OA, widely known as the author-pays model, relies on the use of article processing charges (APCs), where publishers charge a fee for publishing their articles OA (Borrego, 2023). As the OA publishing landscape continues to experience large-scale growth (Piwowar et al., 2018; Shu & Larivière, 2024), the APC model is evolving in tandem with the support of funder and institutional OA policies and, more recently, through read-and-publish and so-called "transformative" agreements (Borrego et al., 2020).

The author-pays model has increasingly become a popular revenue source for publishers, either in addition to the subscription model or as the sole income stream (Crawford, 2024; Suber, 2012). While the "Big 5" commercial publishers – Elsevier, Sage, Springer Nature, Taylor & Francis, and Wiley – use APCs to supplement existing revenue strategies (Butler, 2023; Butler et al., 2023), other publishers, such as Frontiers and MDPI, rely solely on APCs to generate revenue (Rodrigues et al., 2020). APCs have proven lucrative and profitable for publishers (Butler et al., 2023; Shu & Larivière, 2024), but highly controversial among researchers who struggle to pay several thousand dollars per article or find they divert resources away from their research and act as a major barrier to publishing OA (Halevi & Walsh, 2021; Nicholas et al., 2024). Against this background, the diamond OA model has recently been favored by policy makers and researchers as neither authors nor readers are excluded on economic grounds. While diamond OA represents community-driven and academic-led and owned publishing initiatives (Global Summit on Diamond Open Access, 2023), ensuring the financial sustainability of diamond OA journals is an ongoing challenge (Becerril et al., 2021; Bosman et al., 2021; Yoon et al., 2024).

Given the prevalence of APC-based models, there is a need for reliable data that can support evidence-based decision making by institutions, funders, or consortia during negotiations with publishers or in the context of science policy making. The lack of transparency around the payment of APCs, combined with the decentralization of individual authors paying from their own funds, has limited the ability of most institutions and the global community at large to know how much is spent on OA publication fees. Studies analyzing APCs at the funder, country, institution, or disciplinary level have encountered limitations. For example, without contacting individual authors, it is difficult to identify who was responsible for paying the APC or whether waivers or discounts were in place (Shamash, 2016; Zhang et al., 2022). The commonly used APC data from the Directory of Open Access Journals (DOAJ) does not include hybrid journals and lacks historical fees (Asai, 2023). While institutions are better positioned to track their own APC expenditure, such information is typically not systematically tracked, leaving institutions to employ a range of methods to estimate their affiliated authors' spend. These estimates are not often publicly available. A notable exception is the OpenAPC initiative in Germany (Pieper & Broschinski, 2018), which is a unique dataset on APCs actually paid by universities, funders, and research institutions from Europe and North America. To date, OpenAPC provides APC records paid for 224,962 articles with a total amount of almost €448 million by 429 institutions from 2005 to 2024 (OpenAPC, n.d.).

As a result, studies have employed a variety of methods to collate APCs and fill in missing data points, such as using an average APC (Schimmer et al., 2015; Smith et al., 2016; Swan & Houghton, 2012), applying the most current APC fee for a journal to older publications (Kendall, 2024; Zhang et al., 2022), or relying on self-reported data via the OpenAPC initiative (Schönfelder, 2020). These approaches may serve the objectives of their studies, but risk significant over- or under-estimating.

A different approach was taken by Butler et al. (2022, 2023), who collected annual APCs for 6,252 journals from several open datasets and historic journal website snapshots via Wayback Machine. To complement and expand upon this previous work, which was limited to journals published by the Big 5 and the 2015 to 2018 period, the present study is based on a new dataset (Butler et al., 2024b) that was created using a similar approach, but includes annual list prices for six large publishers (Elsevier, Frontiers, MDPI, PLOS, Springer Nature, and Wiley) and spans 2019 to 2023 (Butler et al., 2024a). We use this new dataset to estimate the global amount of APCs paid for OA publishing to these six publishers in recent years. More specifically, we seek to address the following research questions:

1.  How much was paid in APCs to these six publishers for the 2019–2023 period?
    a.  How did the estimated spend differ between gold and hybrid OA?
    b.  How did the estimated spend differ between publishers?
    c.  How did the estimated spend develop over time?
2.  How do gold and hybrid APCs paid (article level) compare to the APCs listed (journal level)?

# 2. Methods

This study largely follows the methodology of estimating APC spend outlined in detail in Butler et al. (2023). We obtained annual APC data from a new open dataset (Butler et al., 2024b) and the number of publications per journal per year from OpenAlex.

## 2.1 Annual APCs

We compiled a new open dataset (Butler et al., 2024b) which combines and standardizes data from the APC price lists of six large publishers (namely, Elsevier, Frontiers, MDPI, PLOS, Springer Nature, and Wiley). The dataset includes APC prices for 8,712 unique journals and 36,618 data points (i.e., journal-year combinations) spanning five years (2019-2023). The APC data was imported from 37 publisher price lists and cleaned to produce one coherent and reusable dataset. A detailed description of the dataset is given in Butler et al. (2024a).

Most publisher price lists include at least one ISSN, but do not always specify whether the identifier provided corresponds to the print (pISSN) or electronic (eISSN) serial number. Therefore, the dataset does not make such a distinction and instead has two ISSN columns (ISSN_1 and ISSN_2). In cases where publishers provide a single ISSN, we assign it to ISSN_1; in cases where publishers distinguish between eISSN and a pISSN, we assign the eISSN to ISSN_1 and the pISSN to ISSN_2.

Publishers provided APCs in multiple currencies including USD, EUR, GBP, JPY and CHF. USD was the most frequently provided currency at 92% of the 36,618 journal-year combinations. The remaining 8% without USD original fees were converted from APCs provided in CHF, EUR or GBP using the average annual rate reported by OFX (2024). This study is based on APCs in USD.

Springer Nature price lists between 2021 and 2023 included several journals with so-called rapid service fees (RSFs), which promise accelerated review and production time to decrease publication delays. For example, *Dermatology and Therapy* (unique_id=4797) advertises rapid publication in the form of two weeks for peer review and three to four weeks from acceptance to online publication, charging a fee of $6,850 in 2023. RSFs present an additional layer of complexity for cost estimation since, at least for the Springer Nature journals in our dataset, they seem to either replace APCs (i.e., 10 gold OA journals) or they are charged in addition to OA fees (i.e., 1 hybrid journal). We therefore treated RSFs as an APC for ten gold journals (because RSFs were the only option for publishing in these journals), while we disregarded the RSF for one hybrid journal (since the RSF was optional and in addition to an APC).

## 2.2 Number of OA articles per journal per year

To estimate APC spend, we determined the number of gold or hybrid OA publications per journal per year in June 2024. This was done by querying the relevant journals, articles, and their access status in OpenAlex (Priem et al., 2022). Since OpenAlex is a new and evolving citation index with some known metadata limitations (Alperin et al., 2024; Culbert et al., 2024; Haupka et al., 2024; Zhang et al., 2024), we cross-checked the number of publications per journal per year from Dimensions to compare results and gain greater confidence of our APC spend estimates. Due to its wider and more inclusive indexing criteria, Dimensions was prioritized over other databases (e.g., Web of Science, Scopus), assuming that both OpenAlex and Dimensions would index all journal articles with DOIs from the six publishers analyzed.

### 2.2.1 Relevant document types

A main challenge is determining which articles can be plausibly considered to be "APC-able", that is, which of the documents published would have been subject to an APC. Generally, research and review article types fall into this category, while letters to the editor, corrections, viewpoints, and editorials do not. We identified documents in OpenAlex by filtering for those classified as *article* or *review*[1], with a publication date between 2019 and 2023, and matching their ISSN to our price list dataset (either ISSN_1 or ISSN_2). The same approach was taken in Dimensions, querying for documents classified as *research article* or *review article*. Publication counts for each journal and year showed good agreement and total spend calculation was within 2% between the two sources, leading us to confidently report estimates based on OpenAlex data.

---

[1] OurResearch introduced four new work types (including *review*) in OpenAlex in May 2024, allowing for a more granular analysis of document types. See OpenAlex release note from 30 May 2024:
https://github.com/ourresearch/openalex-guts/blob/dcd1627b75152c0127b610dc82217f6059e3a879/files-for-datadumps/standard-format/RELEASE_NOTES.txt#L12

## 2.2.2 Open access status

Every journal-year combination in our dataset was classified as either gold or hybrid in the respective publisher price list. Gold journals are those for which all documents are APC-able OA, while hybrid journals are subscription-based and offer OA for individual articles upon payment of an APC. For gold journals, we did not use the OA status returned by OpenAlex due to known fluctuations and inaccuracies of the underlying Unpaywall algorithm (Jahn et al., 2021; Sanford, 2022; Schares, 2023). Instead, we considered all documents published in these journals as APC-able OA. Among the 9,494 gold journal-year combinations in the dataset, 8,380 reported APCs above $0. Of these, 8,221 combinations could be matched to at least one article when matching by ISSN in OpenAlex. This matching in OpenAlex returned 2,591,747 gold OA articles published between 2019 and 2023 that we assume have been subjected to an APC.

For journals listed as hybrid, there were 27,113 journal-year combinations in the APC dataset, of which 26,994 contained APC information. When matching with OpenAlex using the journals' ISSN, we were able to identify at least one article for 26,554 journal-year combinations. In these journals, we only considered the portion of articles available as hybrid OA as having been subjected to an APC. We therefore relied on the OpenAlex OA status of each article to determine how many articles were APC-able.

When analyzing OA status of articles in hybrid journals, we noticed that some journals returned very high rates of hybrid OA (sometimes as high as 100%) which contradicts reports of much lower hybrid rates around 4-10% (Jahn, 2024; Jahn et al., 2022; Piwowar et al., 2019). Our investigation revealed that many of these corresponded to journals with delayed OA (i.e., subscription journals that make all their content available OA after a journal-specific embargo period). For example, we identified 154 journals that belong to Elsevier's Open Archive program (Elsevier, 2024). Some articles in these journals were inaccurately classified by OpenAlex as hybrid (Jahn et al., 2021; Schares, 2023). To correct these misclassifications, we only considered APC-able articles for which OpenAlex was able to identify a Creative Commons (CC) license (i.e., CC-BY, CC-BY-NC, CC-BY-NC-ND). OurResearch has noted they are making a similar change to the OA status classification in Unpaywall and OpenAlex. As shown in Figure 1 for the Elsevier delayed OA journal *Neuron*, delayed OA articles can be detected by their "publisher-specific-oa" license, while CC-BY licenses are assumed to represent articles for which a hybrid OA APC was paid.

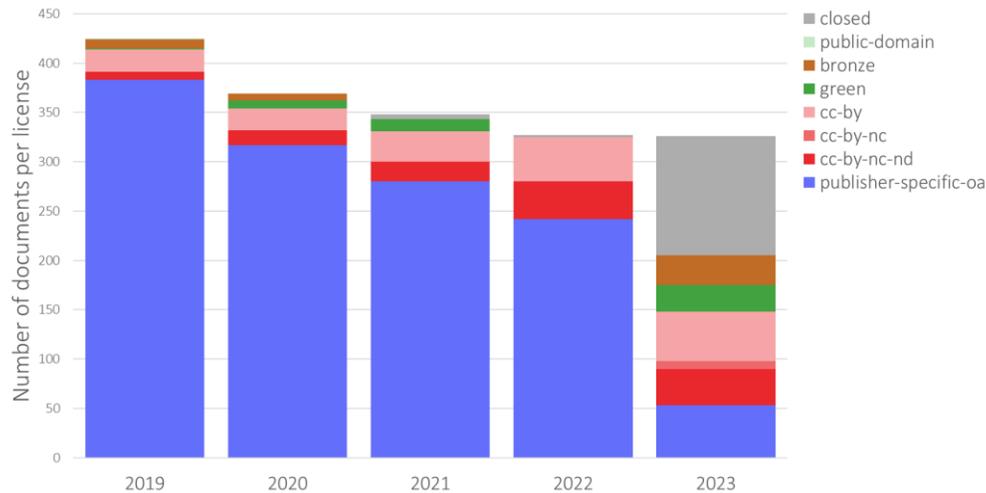

**Figure 1**. Example of license distribution for Elsevier delayed OA journal *Neuron* at time of analysis. CC-license variants are shown in pink/red colors.

After limiting hybrid articles to those with a CC-license variant from all publishers, the number of APC-able hybrid OA articles in our overall dataset decreased by 68.4%. The resulting dataset contained 24,747 journal-year combinations with at least one APC-able article for 635,066 articles published in hybrid OA journals between 2019 and 2023 that we assume have been subjected to an APC.

## 2.3 Accounting for inflation

When reporting monetary trends over time it is useful to account for inflation. Inflation rates were particularly high in the five-year period 2019 to 2023. For example, the global CPI was 25.4%, the CPI for the US was 19.1% and the CPI for Advanced Economies was 16.5% (International Monetary Fund, 2024). We took the view that the CPI Advanced Economies is the most suitable indicator for a general inflation adjustment for APCs reported by the six publishers involved.

Inflation adjustment was carried out on the list prices, adjusting each annual APC to 2023 dollars using the CPI Advanced Economies, as reported by the International Monetary Fund (2024). For example, since the CPI Advanced Economies from 2019 to 2023 was 1.165, the nominal 2019 APC of $3,000 was converted to $3,496 to obtain a 2023 equivalent value. Ultimately, since the academic journal market is an international endeavor with many parties across the globe involved in the production, any more nuanced inflation adjustment could be subject to criticism. Readers wishing to view comparisons using a different CPI are invited to do so using the data provided in Butler et al. (2024b).

When reporting our results, we present both the inflation adjusted as well as nominal amounts. Inflation adjusted numbers are helpful to demonstrate trends in annual prices, while the nominal amounts represent actual fees paid, enabling libraries and funders to make comparisons with their annual budgets.

## 2.4 Calculating global APC spend

The number of APC-able documents per journal per year was multiplied with the annual APC data. For the 13 out of the 35,508 data points for which we had two APC values per year (either due to change of OA status or a change in publisher), we matched all documents for that journal and year to the first APC value (APC_order=1), assuming that papers were more likely to be asked to pay the first value. For example, for the journal *Alzheimer's & Dementia: Diagnosis, Assessment & Disease Monitoring* (unique_id=107) for which we had an APC of $2,000 from Elsevier (APC_order=1) and an APC of $2,200 from Wiley (APC_order=2) in 2020, we assigned all documents published in 2020 to Elsevier using the $2,000 price point (documents for 2023 would be matched to Wiley at the $2,200 price point). Finally, summary statistics such as sum, mean, and median were calculated for various aggregates such as by publisher, year and OA status. These results are presented below.

## 2.5 Limitations

Our approach is limited in several regards. First and foremost, due to the lack of transparent information about the amount of actual OA fees paid, we are *estimating* global spend based on APC list prices and documents published. Only publishers know the exact revenue they generate from gold and hybrid OA.

**Table 1.** Journal-year combinations treated as 100% waivers.

| Publisher | Number of journal-year combinations | | Number of journal-year combinations treated as 100% waivers | | |
|---|---|---|---|---|---|
| | total | APC >$0 | APC=$0 | no APC reported | % of total |
| Elsevier | 12,478 | 12,113 | 0 | 365 | 2.9% |
| Frontiers | 618 | 599 | 19 | 0 | 3.1% |
| MDPI | 1,676 | 1,643 | 1 | 32 | 2.0% |
| PLOS | 50 | 50 | 0 | 0 | 0.0% |
| Springer Nature | 13,422 | 12,708 | 2 | 712 | 5.3% |
| Wiley | 8,374 | 8,272 | 101 | 1 | 1.2% |
| all publishers | 36,618 | 35,385 | 123 | 1,110 | 3.4% |

Publishers or academic societies sometimes offer rebates or waivers for APCs. We consider such waivers only when explicitly noted in the list prices. For example, of the 36,618 journal-year combinations in our dataset, 123 were listed as $0 and an additional 1,110 did not provide an APC.

Manual checks on those not listing an APC revealed that some journals were financed by a society, others granted temporary waivers, and in some cases, the journal directly invoiced authors. Table 1 provides an overview of these journal-level data points that we treated as waivers in our study. Beyond those identified in list prices, it is impossible for us to consider discounts and waivers given limited available information.

Journals sometimes offer individual authors rebates or waivers if they are not able to pay the full APC. We are unable to consider these discounts in our study since they are not captured in publicly available metadata. For example, Elsevier, Springer Nature and Wiley provide waivers and discounts to authors from low- to middle-income countries, but without publicly accessible data, we cannot consider these in our analysis. MDPI and Frontiers provide some information on discounts and waivers, but they report very general information that is not sufficient for us to calculate amounts of APCs waived or discounted. For example, MDPI (n.d.) indicates that their discounts range "from 15% in our most established journals, up to 100% in our new or humanities journals" (MDPI, n.d., para. 8). Frontiers reported that they "provided $22.9M in fee support to over 30,000 authors. Of those articles, 44% of authors were from South America, and 38% from African nations" (Frontiers, n.d., para. 19). This amounts to 6.9% of our spend estimate for Frontiers in 2022. However, it is unclear whether this already factors in the 19 journals for which Frontiers had temporarily waived APCs (Table 1). Waiver programs have also been criticized for failing to provide an equitable solution to the author pays model (Rouhi et al., 2022). Overall, we acknowledge that waivers were likely granted to a small fraction of gold OA articles, and possibly to a minute portion of hybrid OA articles.

Finally, our spend estimates also do not account for read-and-publish agreements (Borrego et al., 2020; ESAC, n.d.; Farley et al., 2021). Our calculations might therefore overestimate *fees paid by individual authors*. However, since these agreements are typically calculated based on volume of publications, they are tantamount to *fees paid up-front by libraries*. As such, where such agreements were in place, we argue that our estimate of global spend still reflects the total amount of OA *fees paid by the community*.

# 3. Results

We estimate that globally the amount of APCs paid to the six publishers included in this study was $8.349 billion over the five-year period analyzed, or $8.968 billion adjusted for inflation to 2023 prices (Table 2). Using inflation adjusted estimates, the spend on OA fees almost tripled from $910.3 million in 2019 to 2.538 billion in 2023. Below we report results comparing estimated spend per OA type, publisher, and listed versus paid APCs and journals.

**Table 2.** Estimate of annual APC spend (in million USD) per OA type for actual APCs paid (nominal) and adjusted for inflation using CPI Advanced Economies.

| Publication year | Spend estimate based on APCs in USD (nominal) | | Spend estimate based on APCs in USD (adjusted to 2023 prices) | |
|---|---|---|---|---|
| | Gold | Hybrid | Gold | Hybrid |
| 2019 | 578.5 | 202.7 | 674.1 | 236.2 |
| 2020 | 844.7 | 312.9 | 977.4 | 362.0 |
| 2021 | 1,289.9 | 415.0 | 1,447.7 | 465.8 |
| 2022 | 1,625.3 | 541.6 | 1,700.1 | 566.5 |
| 2023 | 1,766.6 | 771.7 | 1,766.6 | 771.7 |
| 2019-2023 | 6,150.0 | 2,243.9 | 6,565.9 | 2,402.3 |

## 3.1 OA type

Using inflation adjusted estimates, 73.2% of the spend went to gold OA, which can be explained by three of the six publishers only offering gold OA and MDPI and Frontiers publishing at high volume. Gold OA spend increased by 162.1% from $674.1 million in 2019 to 1.767 billion in 2023 and hybrid by 226.8% from $236.2 to $771.7 million (Table 2, Table 3).

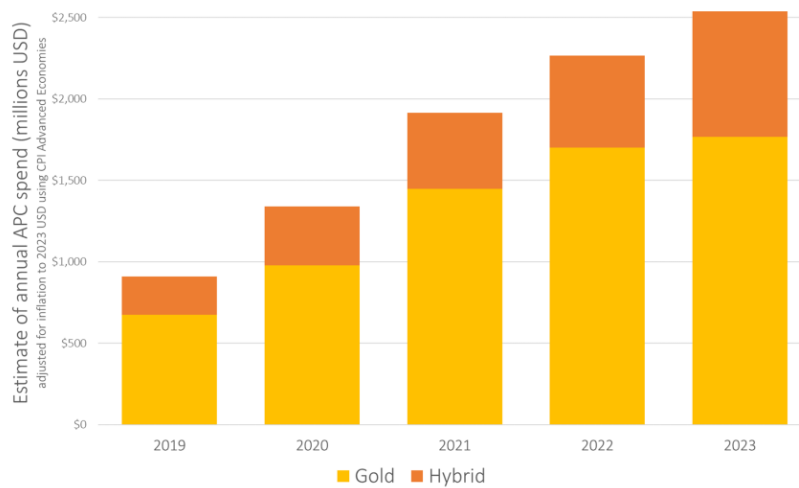

**Figure 2.** Estimate of annual APC spend (in millions USD) per OA type, adjusted for inflation to 2023 USD using CPI Advanced Economies.

## 3.2 Publishers

Analyzing growth in annual APC spend by publisher (Table 3, Figure 3), Elsevier (265.0%), Wiley (263.7%) and MDPI (247.2%) showed the largest increases of revenue from OA fees from 2019 to 2023. These publishers increased their APC income by more than 3.5 times, even when inflation is accounted for. PLOS, the publisher with the smallest APC revenue among the six companies analyzed, is also the only publisher showing no significant growth (1.5%) from 2019 to 2023, as their number of papers published remained constant (+0.2%). Even if some annual increases could be observed until 2021 ($45.7 million), PLOS's revenue has been decreasing since 2022 to $40.7 million in 2023.

**Table 3.** Growth of article output and APC spend (adjusted for inflation to 2023 USD using CPI Advanced Economies) from 2019 to 2023 per publisher and OA type.

| 2019 to 2023 growth rate | Number of APC-able publications | | | Spend estimate based on APCs in USD (adjusted to 2023 prices) | | |
|---|---|---|---|---|---|---|
| | Gold+Hybrid | Gold | Hybrid | Gold+Hybrid | Gold | Hybrid |
| all publishers | +155.7% | +142.3% | +215.2% | +178.8% | +162.1% | +226.8% |
| Elsevier | +271.7% | +306.3% | +225.5% | +265.0% | +282.5% | +248.6% |
| Frontiers | +156.9% | +156.9% | n/a | +149.0% | +149.0% | n/a |
| MDPI | +164.9% | +164.9% | n/a | +247.2% | +247.2% | n/a |
| PLOS | +0.2% | +0.2% | n/a | +1.5% | +1.5% | n/a |
| Springer Nature | +66.9% | +37.8% | +165.5% | +86.7% | +50.2% | 179.5% |
| Wiley | +270.0% | +276.9% | +263.4% | +263.7% | +274.7% | +256.4% |

MDPI became the largest publisher of OA based on estimated APC revenue, when it surpassed Springer Nature in 2021 (Figure 3). However, estimated spend on APCs with MDPI plateaued at 687.4 million in 2022. Due to its comparatively small growth rate (66.9%), particularly in the number of gold OA articles published (37.8%), Springer Nature was surpassed by Elsevier in 2023 as the second largest publisher of OA. Wiley showed a similar increase to Elsevier from 2022 to 2023, which can at least partly be explained by the acquisition of Hindawi journals. Frontiers is the only publisher that sees a significant decline in annual APC revenue when it dropped by 25.5% from 2022 ($346.1 million) to 2023 ($257.9 million). For both Frontiers and MDPI the declines in OA revenue are caused by a declining number of papers published from 2022 to 2023 (Frontiers: -30.7%, MDPI -5.1%), which may be linked to scandals around paper mills (Brundy & Thornton, 2024; Van Noorden, 2023), criticism of using special issues as a growth model (Brainard, 2023; Hanson et al., 2024), and the delisting of journals from the Web of Science in March 2023 due to "increasing threats to the integrity of the scholarly record" (Brainard, 2023, p. 1283). Scandals around delisted journals and thousands of retractions caused Wiley to shut down 19 Hindawi journals compromised by paper mills and to stop using the imprint name (Kincaid, 2023; Subbaraman, 2024). However, this has not caused a similar decline in gold OA articles and therefore APC revenue as per our estimate.

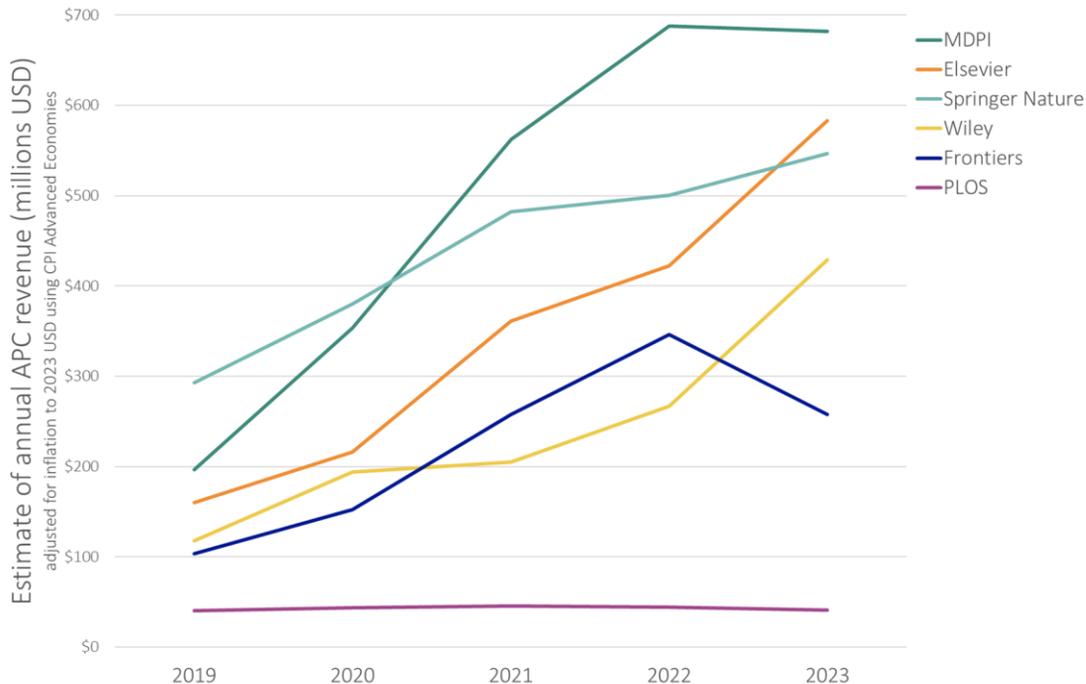

**Figure 3.** Estimate of annual APC revenue (in millions USD) by publisher adjusted for inflation to 2023 USD using CPI Advanced Economies.

As mentioned above and shown in Table 3, spend for hybrid APCs grew faster than that for gold fees due to the faster growth rate in the number of articles published (+215.2% for hybrid versus +142.3% for gold). The growth in the number of hybrid papers may be influenced by read-and-publish agreements, where libraries pay APCs as part of their annual subscription fees. Figure 4 shows the estimated annual spend per publisher per year and type of OA. It also shows that publishers who offer both gold and hybrid routes grew their revenue from hybrid faster than that from gold APCs. Wiley and Elsevier increased hybrid fee revenue from $71.1 and $82.5 million in 2019 to $253.4 and $287.43 million in 2023, representing growth rates of 256.4% and 248.6%, respectively (Table 3). Springer Nature increased their hybrid revenue by 179.5% from 2019 to $230.9 million in 2023. Comparing portfolios of the three publishers that offer both OA routes based on annual estimated revenue, Elsevier seems to focus almost equally on both gold (50.9% of APC revenue) and hybrid (49.1%), while Springer Nature favors gold (62.4%) and Wiley hybrid OA (59.3%). While patterns for Springer Nature and Wiley seem to be a continuation of trends observed for the 2015 to 2018 (Butler et al., 2023), Elsevier is moving to a higher share of gold OA articles from 57.2% in 2019 to 62.5% in 2023.

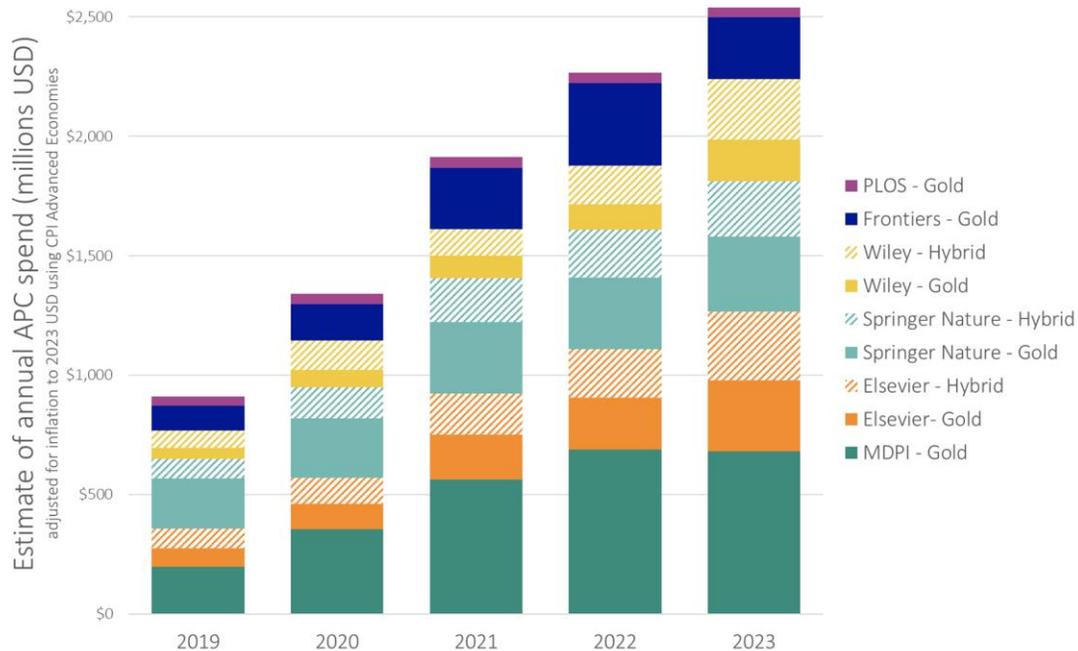

**Figure 4.** Estimate of annual APC revenue (in millions USD) by publisher and OA type adjusted for inflation to 2023 USD using CPI Advanced Economies.

## 3.3 APCs listed vs paid

When analyzing APCs and comparing fees for gold and hybrid OA and across publishers, it is important to highlight the difference between reporting averages or medians of APCs listed, where each price point represents one journal, and APCs paid, where results are computed on the article level. Here we report medians for both listed and paid APCs for 2023 to highlight the difference.

The violin plots in Figure 5 show the distribution of APC prices (blue) and the amount of articles per price that we estimate were actually paid (orange). As has been shown by numerous studies (Björk & Solomon, 2014; Butler et al., 2023; Jahn & Tullney, 2016; Schönfelder, 2020; Smith et al., 2016), the median APCs for gold journals are lower ($2,000) for the six publishers studied than for hybrid ($3,230). However, Figure 5 also reveals that the median APC paid is higher than the median APC listed for both gold ($2,450) and hybrid ($3,600). In other words, authors tend to publish more in journals with higher APCs (or publishers charge higher APCs for popular journals). This trend can be confirmed across publishers (Table 4, Figure 6), except for PLOS, where the median of APCs paid is that of PLOS ONE at $1,805 compared to the median journal-level fee of $2,615. This can be explained by the large volume of papers published in the mega journal PLOS ONE compared to the other eleven PLOS titles.

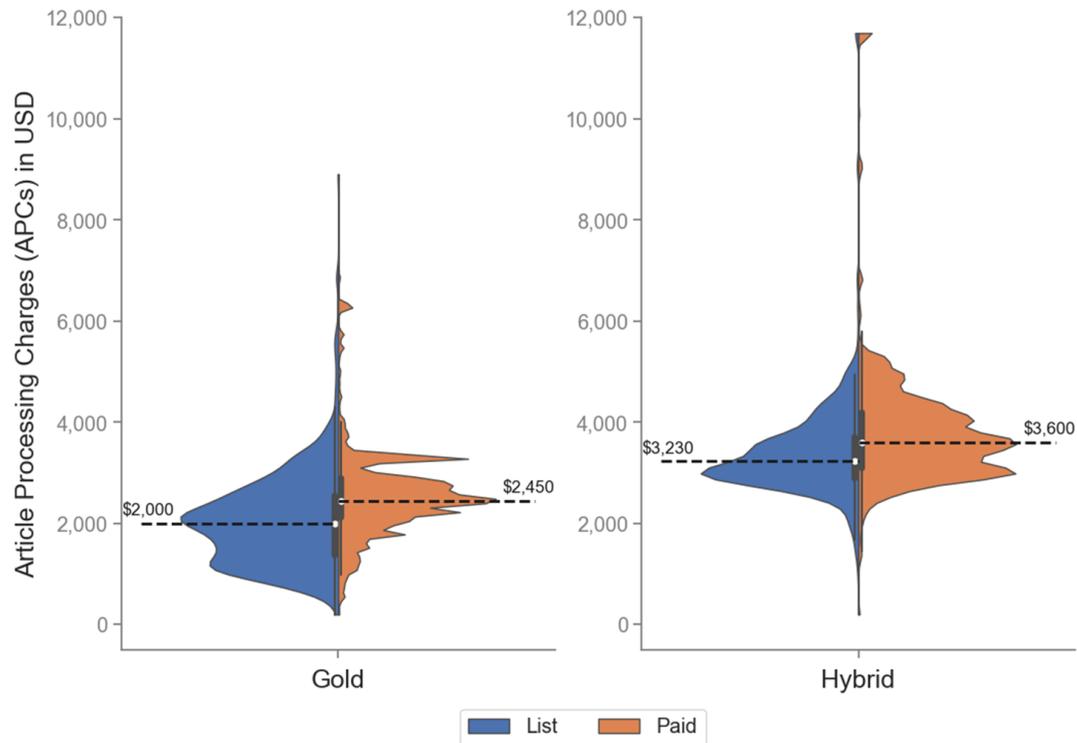

**Figure 5.** Distribution of number of journals/articles per APC amount listed (journal level, in blue) and paid (article level, in orange) for 2023 by OA type. Median values are indicated on a dotted line.

**Table 4.** Median of 2023 APCs listed (journal level) and paid (article level) per OA status per publisher.

|  | Median gold APC 2023 | | Median hybrid APC 2023 | |
| --- | --- | --- | --- | --- |
|  | Listed | Paid | Listed | Paid |
| all publishers | $2,000 | $2,450 | $3,230 | $3,600 |
| Elsevier | $1,800 | $2,100 | $3,220 | $3,480 |
| Frontiers | $2,125 | $3,295 | n/a | |
| MDPI | $1,336 | $2,450 | n/a | |
| PLOS | $2,615 | $1,805 | n/a | |
| Springer Nature | $2,590 | $2,790 | $3,090 | $3,590 |
| Wiley | $2,200 | $2,550 | $3,450 | $4,020 |

Comparing across publishers (Table 4), median list prices for gold titles are highest at PLOS ($2,615) and Springer Nature ($2,590) and lowest at MDPI ($1,336) and Elsevier ($1,800). However, these median list prices do not reflect the APCs most commonly paid. The median APCs paid for gold is highest with Frontiers ($3,295), followed by Springer Nature ($2,795) and lowest with PLOS ($1,805) and Elsevier ($2,100). For the three publishers offering hybrid OA, median listed hybrid APCs were all above $3,000 with Wiley charging $3,450, Elsevier $3,220 and Springer Nature $3,090. The median APCs paid per hybrid article were highest for Wiley at $4,020, followed by Springer Nature ($3,590) and Elsevier ($3,480).

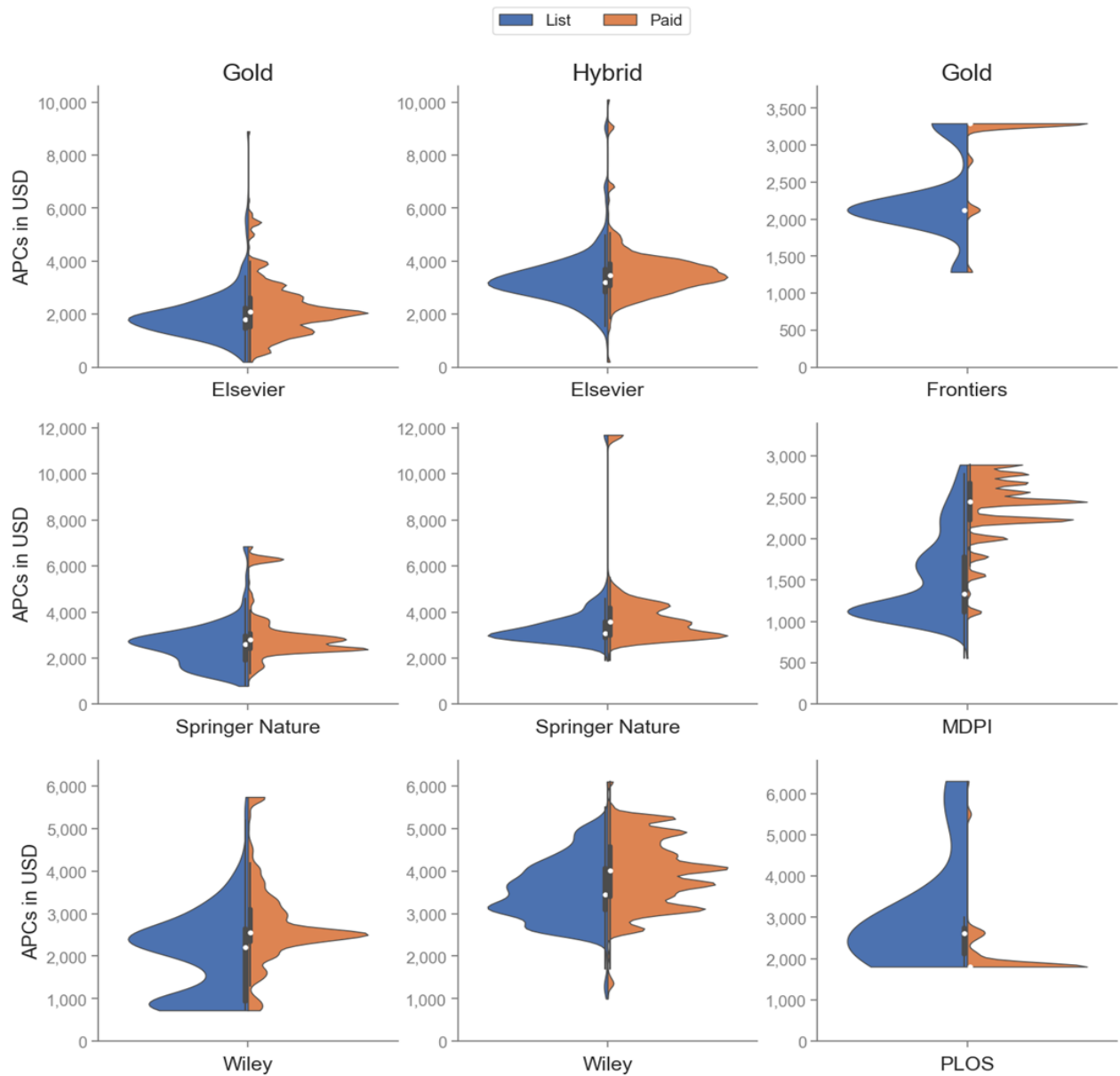

**Figure 6.** Distribution of number of journals and articles per APC price point comparing listed (journal level, in blue) and paid (article level, in orange) data for 2023 by publisher and OA type. Note that y-axes differ between publishers.

A particularly large difference between listed and paid APC can be observed for MDPI and Frontiers, where the higher or highest priced journals were the ones with the highest volume of articles in 2023. This is also emphasized in the detailed violin plots per publisher in Figure 6, where the right side of the violin plots for MDPI and Frontiers are skewed upward towards higher prices. Another interesting pattern in the distribution of paid APCs is the number of articles published in journals with extremely high APCs, in particular at $11,690 for the hybrid option to publish in a *Nature* family journal or $9,080 in a *Cell* title. While Elsevier's most expensive gold titles at $8,900 seem to be less attractive for authors, their *Lancet* gold titles at $5,270 attract more authors. Similarly, a spike in the distribution can also be observed for Springer Nature's and Wiley's higher priced gold OA titles at $6,850 and $5,720, respectively.

# 4. Discussion

Our study shows that the APC model continues to grow and that the combination of a growing volume of gold and hybrid articles and increasing fees leads to global spend almost tripling in five years, even after adjusting for inflation. We find that since 2021, the OA-only publisher MDPI has become the largest publisher of OA in terms of publication output and estimated OA revenue. While MDPI still charges lower APCs, it employs strategies such as special issues to generate high volumes of articles to generate their revenue (Brainard, 2023; Hanson et al., 2024). However, this growth seems to have plateaued in 2022, perhaps due to scandals related to questionable publication practices. Frontiers, which faced similar criticisms around paper mills and AI-generated content, saw a significant drop in the number of papers and therefore APC revenue. At the same time, traditional publishers Elsevier, Wiley and Springer Nature continue to expand their OA offerings but with different pricing strategies across their portfolios, confirming trends reported by Butler et al. (2023). While Springer Nature started out as the largest publisher of OA articles in 2019, they were outgrown by both MDPI and Elsevier. Elsevier, along with Wiley, had particularly steep increases in estimated APC revenue from 2022 to 2023. Both publishers increased their revenue from hybrid APCs more than 3.5 times, which may at least partially be due to their increasing number of read-and-publish agreements, which have libraries pay for OA as part of their annual subscriptions (Borrego, 2023).

In observing the growth in APC revenues, we also demonstrate that higher priced journals tend to attract more authors (or, seen another way, that journals that attract more authors have the highest APCs). The relationship between APC price and journal prestige warrants further attention, especially as publishers seem to use prestige, not costs, to justify fee increases. The disconnect between cost and APC price (Grossmann & Brembs, 2021; Khoo, 2019) also explains the well-known but counterintuitive finding that hybrid APCs are higher than gold fees, even if hybrid journals are already fully funded through the reader-pays model (i.e., subscriptions).

Just as importantly, our approach allowed us to capture and document the growth in listed APC prices and in revenues, both of which were substantial over the five years studied. These yearly fluctuations bring into focus the extent to which studies that use a "back-of-the-napkin" approach overestimate past spending (Kendall, 2024). In contrast, by using a dataset with journal-level data over time, our approach can be used for detailed analysis of institutional or national spending on OA publishing without significantly overestimating previous spending with current APC rates.

We have attempted to overcome the inaccuracies of previous studies by accounting for price fluctuations over time and by linking the number of APC-able articles published per journal per year. The result was a robust and transparent set of calculations, within the limits of what is publicly available. Where possible, we have made methodological choices that would underestimate the total amount spent on APCs worldwide, as any overestimation may mislead libraries and consortia into believing that read-and-publish agreements offer greater savings than they actually do. Even so, our best and most conservative estimates suggest that, by 2023, the world was spending over $2.5 billion dollars in APCs to the six publishers in our study.

# 5. Conclusion and outlook

We believe estimates such as ours can support more informed, data-based decision making by libraries, consortia, funders, and researchers. However, we acknowledge that producing such an aggregate APC dataset with substantial coverage over time was resource intensive and will be costly to extend to other publishers or future time periods. Yet, this information is crucial for the academic community to responsibly make decisions about where to publish, and on how to invest into reasonable mechanisms and approaches for OA. Estimates such as ours are only necessary because of the lack of transparency from publishers. We therefore call upon publishers to be transparent about their OA revenues and to inform the community about the volume and sources of funds for various fees and models, including APCs, read-and-publish agreements, rapid service fees, article development charges, or any other fees paid by academic societies, institutions or governments.

Our study contributes to very lively debates about the APC model, in particular its inequalities, unsustainability and disconnect from the actual cost of OA publishing. While these debates are often heated and based on opinions, our study contributes an evidence base that can help to inform the community about price developments and publishing trends. Even if our data can only be used for estimates and does not capture waivers and discounts, our analysis demonstrates that there is a massive amount of money spent on APCs and that this amount is growing at a rate that is almost certainly unsustainable. The estimates are also enough to indicate the disconnect between APC pricing and cost of publishing. This discrepancy suggests that OA publishing could be achieved at lower costs (e.g., through diamond OA), freeing up research budgets.

To continue contributing to this conversation, we plan to further refine our methodology on obtaining the number of "APC-able" documents by triangulating annual publication counts among OpenAlex, Dimensions and Web of Science. We will also work to fill gaps in the dataset and investigate external funding sources of journals that do not charge APCs to differentiate between temporary waivers and other models (e.g., subscribe to open and diamond OA) (Simard et al., 2024). Future work may extend the analysis to include author affiliations to understand who pays APCs, as well as to investigate the effect of national OA policies to who pays APCs.

## Data availability

The APC dataset analyzed in this paper is available on the Harvard Dataverse under a CC-0 license. It is cited as Butler et al. (2024a) in the reference list and available at https://doi.org/10.7910/DVN/CR1MMV. Inflation adjusted APCs and publication data from OpenAlex will be made available with a future version of this preprint.

## Acknowledgements

We would like to thank Jason Portenoy and Kyle Demes at OurResearch for resolving issues about OA status and document types and for providing us with the publication data and OA status for the journals in our dataset. We also thank Jeffrey Brainard, Diego Kozlowski, Martin Reinhart and Ted Bergstrom for fruitful discussions about inflation adjustments.

## Author contributions

S.H.: Conceptualization, Data curation, Formal analysis, Investigation, Methodology, Project administration, Resources, Supervision, Validation, Visualization, Writing—original draft, Writing—review & editing. E.S.: Conceptualization, Data curation, Formal analysis, Investigation, Methodology, Validation, Visualization, Writing—original draft, Writing—review & editing; J-P.A.: Conceptualization, Data curation, Formal analysis, Investigation, Methodology, Validation, Visualization, Writing—review & editing; M.H.: Data curation, Writing—original draft, Writing—review & editing; L-A.B.: Conceptualization, Data curation, Writing—review & editing; N.S.: Data curation, Methodology, Writing—review & editing

## Competing interests

The authors have no competing interests to declare.

## Funding information

The authors did not receive funding for this project.